\documentclass [a4paper,10pt]{article} 
\usepackage[ascii]{inputenc}
\usepackage{mathtools}
\usepackage{url}
\usepackage{amsfonts}

\title{Everettian probabilities, the Deutsch-Wallace theorem and the Principal Principle}

\author{ Harvey R. Brown\\
Faculty of Philosophy, University of Oxford\\ Radcliffe Observatory Quarter 555\\ 
Woodstock Road, Oxford OX2 6GG, U.K.\\{\em harvey.brown@philosophy.ox.ac.uk}\\
\medskip\\
Gal Ben Porath\\
Department of History and Philosophy of Science\\
University of Pittsburgh\\
1101 Cathedral of Learning,  
4200 Fifth Avenue\\
Pittsburgh, PA USA 15260\\
{\em gab80@pitt.edu}}

\date{}

\begin{document}

\maketitle

\medskip

\textit{Chance, when strictly examined, is a mere negative word, and means not any real power which has anywhere a being in nature.} David Hume (Hume, 2008)

\medskip

\textit{[The Deutsch-Wallace theorem] permits what philosophy would hitherto have regarded as a formal impossibility, akin to deriving an ought from an is, namely deriving a probability statement from a factual statement. This could be called deriving a tends to from a does.} 
                                                                                David Deutsch (Deutsch, 1999)

\medskip

\textit{[The Deutsch-Wallace theorem] is a landmark in decision theory. Nothing comparable has been achieved in any chance theory. \ldots [It] is little short of a philosophical sensation \ldots it shows why credences should conform to [quantum chances].} 
                         Simon Saunders (Saunders, `The Everett interpretation: probability' [unpublished manuscript])

\medskip

\begin{abstract}
This paper is concerned with the nature of probability in physics, and in quantum mechanics in particular. It starts with a brief discussion of the evolution of Itamar Pitowsky's thinking about probability in quantum theory from 1994 to 2008, and the role of Gleason's 1957 theorem in his derivation of the Born Rule. Pitowsky's defence of probability therein as a logic of partial belief leads us into a broader discussion of probability in physics, in which the existence of objective ``chances'' is questioned, and the status of David Lewis' influential Principal Principle is critically examined. This is followed by a sketch of the work by David Deutsch and David Wallace which resulted in the Deutsch-Wallace (DW) theorem in Everettian quantum mechanics. It is noteworthy that the authors of this important decision-theoretic derivation of the Born Rule have different views concerning the meaning of probability. The theorem, which was the subject of a 2007 critique by Meir Hemmo and Pitowsky, is critically examined, along with recent related work by John Earman. Here our main argument is that the DW theorem does not provide a justification of the Principal Principle, contrary to the claims by Wallace and Simon Saunders. A final section analyses recent claims to the effect that the DW theorem is redundant, a conclusion that seems to be reinforced by consideration of probabilities in ``deviant'' branches of the Everettian multiverse.
\end{abstract}

\tableofcontents{}

\section{Quantum probability and Gleason's Theorem} 

Itamar Pitowsky, in attempting to explain why the development of quantum mechanics was a revolution in physics, argued in 1994 that at the deepest level, the reason has to do with probability. 
\begin{quote}
\ldots the difference between classical and quantum phenomena is that relative frequencies of microscopic events, which are measured on distinct samples, often systematically violate some of Boole's conditions of possible experience.\footnote{Pitowsky, 1994, p. 108.}
\end{quote}
Pitowsky's insightful exploration of the connections between's George Boole's 1862 work on probability, polytopes in geometry, and the Bell inequality in quantum mechanics is well known. In 1994, he regarded the experimental violation of the Bell inequality as ``the edge of a logical contradiction''. After all, weren't Boole's conditions of possible experience what any rational thinker would come up with who was concerned about the undeniable practical connection between probability and finite frequencies?

Not quite. Pitowsky was clear that a strict logical inconsistency only comes about if frequencies violating the Boolean conditions are taken from a single sample. Frequencies taken from a batch of samples (as is the case with Bell-type experiments)  need not, for a number of reasons. The trouble for Pitowsky in 1994 was that none of the explanations for the Bell violations that had been advanced until then in quantum theory (such as Fine's prism model or non-local hidden variables) seemed attractive to him. 

This strikes us as more a problem, in so far as it is one, in physics than logic. At any rate, it is noteworthy that in a 2006 paper, Pitowsky came to view the violation of Bell-type inequalities with much more equanimity, and in particular less concern about lurking logical contradictions. Now it is a ``purely probabilistic effect'' that is intelligible once the meaning of probability is correctly spelt out in the correct axiomatisation of quantum mechanics. What is this notion of probability?

 In his 1994 paper, Pitowsky did not define it. He stated that Boole's analysis did not require adherence to any statement about what probability means; it is enough to accept that in the case of repeatable (exchangeable or independent) events, probability is \emph{manifested} in, but not necessarily defined by, frequency. However in 2006, he was more committal:
 \begin{quote}
\dots a theory of probability is a theory of inference, and as such, a guide to the formulation of rational expectations.\footnote{Pitowsky, 2006, p.4.}
\end{quote}
In the same paper, Pitowsky argued that quantum mechanics is essentially a new theory of probability, in which the Hilbert space formalism is a ``logic of partial belief'' in the sense of Frank Ramsey.\footnote{\emph{Ibid}.}
 Pitowsky now questioned Boole's view of probabilities as weighted averages of truth values, which leads to the erroneous ``metaphysical assumption'' that incompatible (in the quantum sense) propositions have simultaneous truth values. Whereas the 1994 paper expressed puzzlement over the quantum violations of Boole's conditions of possible experience by way of experimental violations of Bell-type inequalities, the 2006 paper proffered a resolution of this and other ``paradoxes'' of quantum mechanics.\footnote{We will not discuss here Pitowsky's 2006 resolution of the familiar measurement problem, other than to say that it relies heavily on his view that the quantum state is nothing more than a device for the bookkeeping of probabilities, and that it implies that ``we cannot consistently maintain that the proposition 'the [Schr\"{o}dinger] cat is alive' has a truth value. \emph{Op. cit}. p. 28. For a recent defence of the view that the quantum state is ontic, and not just a bookkeeping device, see (Brown, 2019).}

We happen to have sympathy for Pitowsky's view on the meaning of probability, and will return to it below. For the moment we are interested rather in Pitowsky's 2008 derivation of the Born Rule in quantum mechanics by way of the celebrated 1957 theorem of Gleason (Gleason, 1957).

Gleason showed that the a measure function that is defined over the closed subspaces of a separable Hilbert space $\cal{H}$ with dimension $D \geq 3$ takes the form of the trace of the product of two operators, one being the orthogonal projection on the subspace in question, the other being a semidefinite trace class operator. In Pitowsky's 2006 axiomatisation of quantum mechanics, the closed subspaces of the Hilbert space representing a system correspond to ``events, or possible events, or possible outcomes of experiments''. If the trace class operator operator in Gleason's theorem is read as the statistical (density) operator representing the state of the system, then probabilities of events are precisely those given by the familiar Born Rule. But there is a key condition in reaching this conclusion: such probabilities are \emph{a priori} ``non-contextual''. That, of course, is the rub. As Pitowsky himself admitted, it is natural to ask why the probability assigned to the outcome of a measurement $B$ should be the same whether the measurement is simultaneous with $A$ or $C$, when $A$ and $C$ are incompatible but each is compatible with $B$. 

Pitowsky's argument for probabilistic non-contextualism hinges on the commitment to ``a Ramsey type logic of partial belief'' while representing the event structure in quantum mechanics in terms of the lattice of closed subspaces of Hilbert space. A certain identity involving closed subspaces is invoked, which when interpreted in terms of ``measurement events'', implies that the event $\{B = b_j\}$ ($b_j$ being the outcome of a $B$ measurement) is defined independently of how the measurement process is set up (specifically, of which observables compatible with $B$ are being measured simultaneously).\footnote{The identity is
\begin{equation}
\bigcup_{i = 1}^{k} \left( \{B=b_j\}\cap \{ A=a_i\} \right) = \{ B=b_j\} = \bigcup_{i = 1}^{l} \left( \{B=b_j\}\cap \{ C=c_i\} \right)
\end{equation}
where $A$, $B$ and $C$ measurements have possible outcomes $a_1, a_2, \ldots, a_k$; $b_1,b_2, \ldots, b_r$ and $c_1,c_2, \ldots, c_l$, respectively.} Pitowsky then invokes the rule that \emph{identical events always have the same probability}, and the non-contextualism result follows. (In terms of projection operators, the probability of a projector is independent of which Boolean sublattice it belongs to.)

The argument is, to us, unconvincing. Pitowsky accepted that the non-contextualism issue is an empirical one, but seems to have resolved it by fiat. (It is noteworthy that in his opinion it is by  \emph{not  committing to the lattice of subspaces as the event structure} that advocates of the (Everett) many worlds interpretation require  a separate justification of probabilistic non-contextualism.\footnote{\emph{Op. cit.} footnote 2.})  Is making such a commitment any more natural in this context than was, say, Kochen and Specker's ill-fated presumption (Kochen and Specker 1967) that truth values (probabilities 0 or 1) associated with propositions regarding the values of observables in a deterministic hidden variable theory must be non-contextual?\footnote{For reasons to be skeptical \emph{ab initio} about non-contextualism in hidden variable theories, see (Bell, 1966).} Arguably, the fact that Pitowsky's interpretation of quantum mechanics makes no appeal to such theories -- which must assign contextual values under pain of contradiction with the Hilbert space structure of quantum states -- may weaken the force of this question. We shall return to this point in section 4.3 below when discussing the extent to which probabilistic non-contextualism avoids axiomatic status in the Deutsch-Wallace theorem in the Everett picture.

John Earman (Earman, 2018) is one among other commentators who also regard Gleason's theorem as the essential link in quantum  mechanics between subjective probabilities and the Born Rule. But he questions Pitowksy's claim that in the light of Gleason's theorem the event structure dictates the quantum probability rule when he reminds us that Gleason showed only that in the case of $D > 2$, and where $\cal{H}$ is separable when $D= \infty$, the probability measure on the lattice of subspaces is represented uniquely by a density operator on $\cal{H}$ \emph{iff} it is \emph{countably} additive (Earman, 2018). And countable additivity does not hold in all subjective interpretations of probability. (de Finetti, for instance, famously restricted probability to finite additivity.)

In our view, the two main limitations of such Gleason-based derivations of the Born Rule are the assumption of non-contextualism and the awkward fact that the Gleason theorem fails in the case of $D<3$, and thus can have no probabilistic implications for qubits.\footnote{Note that a Gleason-type theorem for systems with  $D\geq 2$ was provided by Busch (Busch, 2003) but for POMs (positive operator-valued measures) rather than PVMs (projection-valued measures). More recent, stronger Gleason-type results are discussed in (Wright and Weigert, 2019).} One might also question whether probabilities in quantum mechanics need be subjective. Indeed, Earman holds a dualist view, involving objective probabilities as well (and which, as we shall see later, have a separate underpinning in quantum mechanics). 

The possibility of contextual probabilities admittedly raises the spectre of a violation of the no-signalling principle, analogously to the way the (compulsory) contextuality in hidden variable theories leads to nonlocality.\footnote{See for example (Brown and Svetlichny, 1990).} It is noteworthy that \emph{assuming} no superluminal signalling and the projection postulate (which Pitowsky adopts) in the case of measurements on entangled systems, Svetlichny showed in 1998 that the probabilities must be non-contextual and used Gleason's theorem to infer the Born Rule.\footnote{See Svetlichny, 1998. For more recent derivations of the Born Rule based on no-signalling, see Barnum, 2003, and McQueen and Vaidman, 2019. A useful critical review of derivations of the Born Rule -- including those of Deutsch and Wallace (see below) -- is found in Vaidman (2019).}

Prominent Everettians have in recent years defended a derivation of the Born Rule based on decision theoretic principles, one which has gained considerable attention. Critics often see probability as the Achilles' Heel in the many worlds picture,  but for some Everettians (including the authors of the proof) the treatment of probability is a philosophical triumph. It is curious, then, that the two authors, David Deutsch and David Wallace, appear to disagree as to what the result means. The disagreement is rooted in the question of what probability means -- to which we now turn.

\section{The riddle of probability}

\subsection{Chances} 

The meaning of probability in the physical sciences has been a long-standing subject of debate amongst philosophers (and to a lesser extent physicists), and it remains controversial. Let us start with something that we think all discussants of the notion of probability can agree on. That is the existence in Nature of ``chance processes or set-ups'' that lead to more-or-less stable relative frequencies of outcomes over the long term. Such an existence claim is non-trivial, and of a straightforwardly empirical, and hence objective, nature. ``Chances'', the term widely used by philosophers for objective probabilities, are distilled in some way from these frequencies, or at least connected somehow with them (see below). The situation is somewhat reminiscent of Poincar\'{e}'s remark that time in physics is the great simplifier. It is a non-trivial feature of the non-gravitational interactions that choosing the ``right" temporal parameter in the fundamental equations in the (quantum) theory of each interaction -- equations which contain derivatives with respect to this parameter -- results in the greatest simplification of such equations.\footnote{The fact that it is the same parameter in each case makes the phenomenon even more remarkable.} There is a fairly clear sense in which the universal standard metric of time in non-gravitational physics corresponds to the workings of Nature, even if there is nowhere in Nature a system, apart from the Universe itself, that acts as a perfect clock, and even if choosing a non-standard metric does not make physics impossible, just more complicated. But whether time itself is to be reified is a moot point; our own inclination is to deny it. Is objective probability, or chance, any different? Is probability in physics any more of a candidate for reification than time is?

When philosophers write about objective probabilities, or ``chances'', an example that comes up repeatedly is the probability associated with the decay of a radioactive atomic nucleus. Perhaps the most famous example of such an isotope is that of Carbon-14, or $^{14}$C, given its widespread use in anthropology and geology in the dating of objects containing organic material. Its half-life is far better established experimentally than the probability of heads for any given coin (another widely cited example) and the decay process, being quantum, is often regarded as intrinsically stochastic.\footnote{There are good reasons, however, for considering the source of the unpredictability in some if not all classical chance processes as having a quantum origin too; see (Albrecht, 2014). But whether strict indeterminism is actually in play in the quantum realm is far from clear, as we see below.} Donald Gillies has argued that the spontaneity of radioactive decay makes such probabilities ``objective'' rather than ``artifactual'', the latter sort being instantiated by repetitive experiments in physics, which necessarily involve human intervention.\footnote{Gillies, 2000, pp. 177, 178.} Furthermore:
\begin{quote}
Any standard textbook of atomic physics which discusses radioactive elements will give values for their atomic weights, atomic numbers, etc., and also for the probability of their disintegrating in a given time period. These probabilities appear to be objective physical constants like atomic weights, etc., and, like
these last quantities, their values are determined by a combination of theory and experiment. \emph{In determining the values of such probabilities no bets at all are made or even considered.} Yet all competent physicists interested in the matter agree on certain standard values--just as they agree on the values of atomic weights, etc. [our emphasis]\footnote{Gillies, 1972, pp. 150, 151.}
\end{quote}
In a similar vein, Tim Maudlin writes:
\begin{quote}
The half-life of tritium \ldots is about 4499 days. Further experimentation could refine the number \ldots Scientific practice proceeds as if there is a real, objective, physical probability density here, not just \ldots degrees of belief. \emph{The value of the half-life has nothing to do with the existence or otherwise of cognizers.} \ldots\ [our emphasis]\footnote{Maudlin, 2007.}
\end{quote}
David Wallace similarly states that 
\begin{quote}
If probabilities are personal things, reflecting an agent's own preferences and judgements, then it is hard to see how we could be right or wrong about those probabilities, or how they can be measured in the physicist's laboratory. But in scientific contexts at least, both of these seem commonplace. One can erroneously believe a loaded die to be fair (and thus erroneously believe that the probability of it showing `6' is one sixth); \emph{one can measure the cross-section of a reaction or the half-life of an isotope. \ldots}

\ldots as well as the personal probabilities (sometimes called subjective probabilities or credences) there also appear to be objective probabilities (sometimes called chances), which do not vary from agent to agent, and which are the things scientists are talking about when they make statements about the probabilities of reactions and the like. [our emphasis]\footnote{Wallace, 2012, p. 137.} 
\end{quote}
In fact, Wallace goes so far as to say that denial of the existence of objective probabilities is incompatible with a realist stance in the philosophy of science, unless one entertains a ``radical revision of our extant science''.\footnote{\emph{Op. cit.}, p. 138. It should be noted, however, that although Wallace claims that the certain probabilities in statistical mechanics must be objective, he accepts that such a claim is hard to make sense of without bringing in quantum mechanics (see (Wallace, 2013)). For further discussion of Wallace's views on probability in physics, see (Brown, 2017).} At any rate, Wallace is far from unique in denying that such probabilities can be \emph{defined} in terms of frequencies (finite or otherwise)\footnote{For discussion of the problems associated with such a definition, see, e.g., Saunders, 2005, Greaves and Myrvold, 2010, Wallace, 2012, p. 247, and Myrvold, W. C., \emph{Beyond Chance and Credence} [unpublished manuscript], section 3.2.} so it seems to follow that the quantitative element of reality he invokes is distinct from the presumably uncontentious, qualitative one related to stable frequencies, referred to at the beginning of this section.

Many philosophers have tried to elucidate what this element of reality is; the range of ideas, which we will not attempt to itemize here, runs from the more-or-less intuitive notion of ``propensity'' to the abstract ``best system" approach to extracting probabilities from frequencies in the Humean mosaic (or universal landscape of events in physics, past, present and future).\footnote{This last approach is due to David Lewis (Lewis, 1986); an interesting analysis within the Everettian picture of a major difficulty in the best system approach, namely ``undermining'', is found in (Saunders, 'The Everett interpretation: probability' [unpublished manuscript]). It should also be recognised that the best system approach involves a judicious almagam of criteria \emph{we} impose on our description of the world, such as simplicity and strength; the physical laws and probabilities in them are not strict representations of the world but rather our systematization of it. Be that as it may, Lewis stressed that whatever physical chance is, it must satisfy his Principal Principle linking it to subjective probability; see below.} The plethora of interpretations on the part of philosophers has a curious feature: the existence of chance is widely assumed before clarification is achieved as to what it is! And the presumption is based on physics, or what is taken to be the lesson of physics.\footnote{Relatively few philosophers follow Hume and doubt the existence of chances in the physical world; one is Toby Handfield, whose lucid 2012 book on the subject (Handfield, 2012) started out as a defence and turned into a rejection of chances. Another skeptic is Ismael (Ismael, 1996); her recent work (Ismael, 2019) puts emphasis on the kind of weak objectivity mentioned at the start of this section. See also in this connection recent work of Bacciagaluppi, whose view on the meaning of probability is essentially the same as that defended in this paper: (Bacciagaluppi, 2019), section 10.}

\subsection{Carbon-14 and the neutron}

Until relatively recently, the half-life (or 0.693 of the mean lifetime) of $^{14}$C (which by definition is a probabilistic notion\footnote{The half-life is not, \emph{pace} Google, the amount of time needed for a given sample to decay by one half! (What if the number of nuclei is odd?) It is the amount of time needed for any of the nuclei in the sample to have decayed with a probability of 0.5. Of course this latter definition will be expected to closely approximate the former when the sample is large, given the law of large numbers (see below).}) was anomalous from a theoretical point of view. It is many orders of magnitude larger than both the half-lives of isotopes of other light elements undergoing the same decay process, and what standard calculations for nucleon-nucleon interactions in Gamow-Teller beta decay would indicate.\footnote{In 2011, Maris \emph{et al.}  (Maris et al. 2011) showed for the first time, with the aid of a supercomputer, that allowing for nucleon-nucleon-nucleon interactions in a dynamical no-core shell model of beta decay explains the long lifetime of $^{14}$C.}

Measurements of the half-life of $^{14}$C have been going on since 1946, involving a number of laboratories worldwide and two distinct methods. Results up until 1961 are now discarded from the mix; they varied between 4.7K and 7.2K years. The supposedly more accurate technique involving mass-spectrometry has led to the latest estimate of 5700 $\pm$ 30 years (B\'{e} and Chechev, 2012). This has again involved compiling frequencies from different laboratories using a weighted average (the weights depending on estimated systematic errors in the procedures within the various laboratories).

 In the case of the humble neutron, experimental determination of its half-life is, as we write, in a curious state. Again, there are two standard techniques, the so-called `beam' and `bottle' measurements. The latest and most accurate measurement to date using the bottle approach estimates the lifetime as 877.7 $\pm$ 0.7 seconds (Pattie, 2018), but there is a 4 standard deviations disagreement with the half-life for neutron beta decay determined by using the beam technique which is currently not well understood. 
 
 It should be clear that measurements involve frequencies of decays in all these cases.  The procedures might be compared to the measurement of, say, the mass of the neutron. As Maudlin says, further experimentation could refine the numbers.  But for those who (correctly) deny that probabilities can be defined in terms of finite frequencies, the question obviously arises: in what precise sense does taking measurements of frequencies constitute a measurement of probability (in the present case, half-life)? Is Gillies right that nothing like a bet is being made? 
 
  The difference between measuring mass or atomic weights, say, and half-life is not that one case involves complicated, on-going interactions between theory and experiment and the other doesn't. The essential difference is that for the latter, \emph{if it is thought to be objective}, no matter how accurate the decay frequency measurements are and how many runs are taken, the resulting frequencies could in principle all be a fluke, and thereby significantly misleading. That is the nature of chance procedures, whether involving isotopes or coins. Here is de Finetti, taken out of context: 
 \begin{quote}
 It is often thought that these objections [to the claim that probability theory and other exact sciences] may be escaped by making the relations between probabilities and frequencies precise is analogous to the practical impossibility that is encountered in all the experimental sciences of relating exactly the abstract notions of the theory and the empirical realities. The analogy is, in my view illusory \ldots in the calculus of probability it is the theory itself which obliges us to admit the possibility of all frequencies. In the other sciences the uncertainty flows indeed from the imperfect connection between the theory and the facts; in our case, on the contrary, it does not have its origin in this link, but in the body of the theory itself \ldots.\footnote{de Finetti, 1964, p. 117. de Finetti's argument, as Gillies (Gillies, 2000) stresses on pages 103 and 159, was made in the context of the falsifiability of probabilistic predictions, even in the case of subjective probabilities. But the argument also holds in the case of inferring ``objective'' probabilities by way of frequencies.}
 \end{quote}
 
  In practice, of course, given ``enough'' runs, the weighted averages of the frequencies are taken as reliable guides to the ``objective'' half-life, given the (weak) law of large numbers. It would be highly improbable were such averaged frequencies to be significantly different from the ``real'' half-life (assuming we trust our experimental methods).  Note first that this meta-probability is in the nature of a belief, or ``credence'' on the part of the physicist; it is not based on frequencies on pain of an endless regress. So this is a useful reminder that in physics whether or not we need to appeal to the existence of objective chances, we certainly need to appeal in an important way to a subjective, or personal notion of probability, even in quantum mechanics. And objectivists, at least, \emph{are} effectively betting -- that the observed frequencies are ``typical'', so as to arrive at something close to the ``objective'' half-life.\footnote{Saunders puts the point succinctly:
 \begin{quote}
Chance is measured by statistics, and perhaps, among observable quantities, only by statistics, but only with high chance. (Saunders, 2010)
\end{quote}
This point is repeated in (Saunders, 'The Everett interpretation: probability' [unpublished manuscript], section 2). To repeat, this ``high chance'' translates into subjective confidence; Saunders, like many others, believes that chances and subjective probabilities are linked by way of the ``Principal Principle'' (see below). Wallace's 2012 version of the law of large numbers explicitly involves both ``personal'' and objective probabilities. His personal prior probabilities Pr($\cdot|C$) are conditional on the proposition expressing all of his current beliefs, denoted by $C$. $X_p$ denotes the hypothesis that the objective probability of $E_n$ is $p$, where as usual it is assumed that $p$ is an \emph{iid}. Then, Wallace claims, the Principal Principle  ensures that his personal probability of the proposition $Y_i$, that heads occurs in the $i$th repetition of the random process, is $p$: symbolically Pr($Y_i|X_p \& C$) = $p$. Given the \emph{iid} assumption for $p$ and the usual updating rule for personal probabilities, it follows from combinatorics that the personal probability Pr($K_M|X_p \& C$) is very small unless $p= M/N$, where $K_M$ is the proposition that the experiment results in $M$ heads out of $N$. Wallace concludes that 
\begin{quote}
\ldots as more and more experiments are carried out, any agent conforming to the Principal Principle will become more and more confident that the objective probability is close to the observed relativity frequency. 
(Wallace, 2012, p.141.)
\end{quote}

Note that there are versions of the ``objective'' position that avoid the typicality assumption mentioned above, such as that defended in (Gillies, 2000), Chapter 7, and (Greaves and Myrvold, 2010). For more discussion of the latter, see section 2.3 (ii) below.}

  \subsection{The law of large numbers}
 
(i)  Whether we are objectivists, subjectivists or dualists about probability, for such processes as the tossing of a bent coin, or the decay of Carbon-14, we will need to be guided by experience in order to arrive at the relevant probability. For the objectivist, this is letting relative frequencies uncover an initially unknown objective probability or chance. For the subjectivist, it is using the standard rule of updating a (suitably restricted) subjective prior probability in the light of new statistical evidence, subject to a certain constraint on the priors. 
 
 Suppose we have a chance set-up, and we are interested in the probability of a particular outcome (say heads in the case of coin tossing, or decay for the nucleus of an isotope within a specified time) in the $n+1$th repetition of the experiment, when that outcome has occurred $k$ times in the preceding $n$ repetitions. For the objectivist, the chance $p$ at each repetition (``Bernoulli trial'') is typically assumed to be an identically distributed and independent distribution (\emph{iid}), which means that the objective probability in question is $p$, whatever the value of $M$ is. But when $p$ is unknown, we expect frequencies to guide us. It is their version of the (weak) law of large numbers that allows the objectivists to learn from experience.\footnote{This is not to say that the law of large numbers allows for probabilities to be \emph{defined} in terms of frequencies, a claim justly criticised in (Gillies, 1973), pp. 112-116.} 
 
The celebrated physicist Richard Feynman defined the probability of an event as an essentially time-asymmetric notion, \emph{viz}. our estimate of the most likely relative frequency of the event in $N$ future trials.\footnote{Feynman et al. 1965, section 6.1.} This reliance on the role of the estimator already introduces an irreducibly subjective element into the discussion.\footnote{See (Brown, 2011). In so far as agents are brought into the picture, who remember the past and not the future, probability requires the existence of an entropic arrow of time in the cosmos; see in this connection (Handfield, 2012, chapter 11), Myrvold (2016) and (Brown, 2017), section 7.} And note how Feynman describes the experimental determination of probabilities in the case of tossing a coin or a similar ``chancy'' process. 
\begin{quotation}
We have defined $P(H) = \langle N_H \rangle /N$ [where $P(H)$ is the probability of heads, and $\langle N_H\rangle$ is the expected number of heads in $N$ tosses]. How shall we know what to \emph{expect} for $N_H$? In some cases, the best we can do is observe the number of heads obtained in large numbers of tosses. For want of anything better, we must set $\langle N_H \rangle = N_H$(observed). (How could we expect anything else?) We must understand, however, that in such a case a different experiment, or a different observer, might conclude that $P(H)$ was different. We would \emph{expect}, however, that the various answers should agree within the deviation $1/2\sqrt{N}$ [if $P(H)$ is near one-half]. An experimental physicist usually says that an ``experimentally determined'' probability has an ``error'', and writes
\begin{equation}
P(H) = \frac{N_H}{N} \pm \frac{1}{2\sqrt{N}}
\end{equation}
There is an implication in such an expression that there is a ``true'' or ``correct'' probability which \emph{could} be computed if we knew enough, and that the observation may be in ``error'' due to a fluctuation. There is, however, no way to make such thinking logically consistent. It is probably better to realize that the probability concept is in a sense subjective, that it is always based on uncertain knowledge, and that its quantitative evaluation is subject to change as we obtain more information.\footnote{Feynman et al. 1965,  section 6-3. We are grateful to Jeremy Steeger for bringing this passage to our attention. For his careful defence of a  more objective variant of Feynman's notion of probability, see (Steeger, 'The Sufficient Coherence of Quantum States' [unpublished manuscript]).}
\end{quotation}
Now how far this position takes Feynman away from the objectivist stance is perhaps debatable; note that Feynman does not refer here to quantum indeterminacies. But other physicists have openly rejected an objective interpretation of probability; we have in mind examples such as E. T. Jaynes (Jaynes, 1963), Frank Tipler (Tipler, 2014), Don Page (Page, 1995) and Jean Bricmont (Bricmont, 2001). In 1995, Page memorably compared interpreting the unconscious quantum world probabilistically to the myth of animism, i.e. ascribing living properties to inanimate objects.\footnote{See (Page, 1995). Page has also defended the view that truly testable probabilities are limited to conditional probabilities associated with events defined at the same time (and perhaps the same place); see (Page, 1995).  This view is motivated by the fact that strictly speaking we have no direct access to the past or future, but only present records of the past, including of course memories. The point is well-taken, but radical skepticism about the veracity of our records/memories would make science impossible, and as for probabilities concerning future events conditional on the past (or rather our present memories/records of the past) \emph{they will be testable when the time comes}, given the survival of the relevant records/memories. So it is unclear whether recognition of Page's concern should make much difference to standard practice.}

(ii) It is indeed worth reminding ourselves of the fact that the brazen subjectivists can make a good -- though far from uncontroversial -- case for the claim that their practice is consistent with standard experimental procedure in physics.
  
 The subjectivist specifies a prior subjective probability for such a probability (presumably conditional on background knowledge of some kind) and appeals to Bayes' rule for updating probability when new evidence (frequencies) accrues. Following the work principally of de Finetti, the crucial assumption that replaces the \emph{iid} condition for objectivists is the constraint of ``exchangeability'' on the prior probabilities, which opens the door to learning from the past. de Finetti's 1937 representation theorem based on exchangeability leads to an analogue of the Bernoulli law of large numbers.\footnote{de Finetti, 1964; for a review of de Finetti's work see (Galavotti, 1989). A helpful introduction to the representation theorem is given by Gillies (Gillies, 2000, chapter 4); in this book Gillies provides a sustained criticism of the subjectivist account of probabilities.} 
 In the words of Brian Skyrms,
 \begin{quote}
de Finetti \ldots looks at the \emph{role} that chance plays in standard statistical reasoning, and argues that that role can be fulfilled perfectly well without the metaphysical assumption that chances exist. (Skyrms, 1984)
\end{quote}

A detailed discussion of the role of the de Finetti representation theorem in physics was provided in 2010 by Greaves and Myrvold, which appears to question de Finetti's own purely subjectivist interpretation of the theorem. Consider an agent looking for the optimal betting strategy for sequences of a relevant class of experimental setups. Then
\begin{quote}
\ldots when he conditionalizes on the results of elements of the sequence, he learns about what the optimal strategy is, and he is certain that any agent with non-dogmatic priors on which the sequence of experiments is exchangeable will converge to the same optimal strategy. If this is not the same as believing that there are objective chances, then it is something that serves the same purpose. Rather than eliminate the notion of objective chance, we have uncovered, in [the agent's] belief state, implicit beliefs about chances, -- or, at least, about something that plays the same role in his epistemic life. \ldots

\ldots  like it or not, an agent with suitable preferences acts as if she believes that there are objective chances associated with outcomes of the experiments, about which she can learn, provided she is non-dogmatic. \ldots There may be more to be said about the nature and ontological status of such chances, but, whatever more is said, it should not affect the basic picture of confirmation we have sketched.\footnote{Greaves and Myrvold (2010), section 3.3.}
\end{quote}
Note that in explicitly leaving the question of the ``nature and ontological status" of chances open, Greaves and Myrvold's analysis appeals only to the intersubjectivity of the probabilistic conclusions drawn by (non-dogmatic) rational betting agents who learn from experience according to Bayesian updating. Whatever one calls the resulting probabilities, it is questionable whether they are out there in the world independent of the existence of rational agents, and playing the role of elements of reality seemingly demanded by Gillies, Maudlin and Wallace as spelt out in section 2.1 above.\footnote{It is noteworthy that in footnote 4 (\emph{op. cit.}), Greaves and Myrvold entertain the view that ``it is via [the Principal Principle] that we ascribe beliefs about chances to the agent''. This Principle is the topic of the next section of our paper, in which we report Myrvold's more recent claim that it is not about chances at all.}\footnote{Greaves and Wallace also take issue (p. 22) with the claim made in the final sentence of section 2.2 above that objectivists need to introduce the substantive additional assumption that the statistical data is ``typical''. As far as we can see, this is again because the notion of chance they are considering is ontologically neutral.}

(iii) In anticipation of the discussion of the \emph{refutability} of probabilistic assertions in science below (section 6), it is worth pausing briefly to consider an important feature of subjectivism in this sense. A subjective probability for some event, either construed as a prior, or conditional on background information (say relevant past frequencies), is not open to revision. Such a probability is immutable; it is not to be ``repudiated or corrected".\footnote{See (de Finetti, 1964, pp. 146, 147).} The same holds for the view that conditional probabilities are rational inferences arising from incomplete information. What the introduction of relevant new empirical information does is not to correct the initial probability but replace it (by Bayesian updating) with a new one conditional on the new (and past, if any) evidence, which in turn is immutable.\footnote{Gillies (Gillies, 2000, pp. 74, 83, 84) and Wallace (Wallace, 2012 p. 138), point out that de Finetti's scheme of updating by Bayesian conditionalisation yields reasonable results only if the prior probability function is appropriate. If it isn't, no amount of future evidence can yield adequate predictions based on Bayesian updating. This a serious concern, which will not be dealt with here (see in this connection Greaves and Myrvold, 2010). Note that it seems to be quite different from the reason raised by Wallace (see section 2.1 above) as to why the half-life of a radioactive isotope, for example, must be an objective probability. See also section 5.1 below.}

(iv) We note finally that, as Myrvold has recently stressed, there is a tradition dating back to the 19th century of showing another way initial credences can ``converge'', which does not involve the accumulation of new information. The ``method of arbitrary functions'', propounded in particular by Poincar\'{e}, involves dynamical systems in which somewhat arbitrary prior probability distributions are ``swamped'' by the physics, which is to say that owing to the dynamics of the system a wide range of prior distributions converge over time to a final distribution. It is noteworthy that Myrvold accepts that the priors can be epistemic in nature, but regards the final probabilities as ``hybrid'', neither epistemic nor objective chances. He calls them \emph{epistemic chances}.\footnote{Myrvold, W. C., \emph{Beyond Chance and Credence} [unpublished manuscript], chapter 5.} 

 \subsection{The Principal Principle}
 
 If one issue has dominated the philosophical literature on chance in recent decades -- and which will be relevant to our discussion of the Deutsch-Wallace theorem below -- it is the \emph{Principal Principle} (henceforth PP), a catchy term coined and made prominent by David Lewis in 1980 (Lewis, 1980), but referring to a longstanding notion in the literature.\footnote{See (Strevens, 1999). In our view, a refreshing exception to the widespread philosophical interest in the PP is Gillies' excellent 2000 book (Gillies, 2000), which makes no reference to it.}
 
 This has to do with the second side of what Ian Hacking called in 1975 the Janus-faced nature of probability\footnote{Hacking, 1984, though see Gillies (Gillies, 2000, pp. 18, 19) for references to earlier recognition of the dual nature of probability. Note that Wallace has argued that the PP grounds both sides; see footnote  23 above.}, or rather its construal by objectivists. The first side concerns the issue touched on above, i.e. the estimation of probabilities on the basis of empirical frequencies. Papineau (Papineau, 1996) called this the \emph{inferential link}between frequencies and objective probabilities (chances). The second side concerns what we do with these probabilities, or why we find them useful. Recall the famous dictum of Bishop Butler that probabilities are a guide to life. Objective probabilities determine, or should determine, our beliefs, or ``credences'', about the future (and less frequently the past) which if we are rational we will act on. Following Ramsey, it is widely accepted in the literature that this commitment can be operationalised, without too much collateral damage, by way of the act of betting.\footnote{See (Gillies, 2000, chapter 4).}
 
 Papineau (Papineau, 1996) called the connection between putative \emph{objective} probabilities (chances) and credences  the \emph{decision-theoretic link}. More famously, it is expressed by way of the Principal Principle; here is a simple version of it: 
 \begin{quote}
 For any number $x$, a rational agent's personal probability, or credence, of an event $E$ conditional on the objective probability, or chance, of $E$ being $x$, and on any other accessible background information, is also $x$.\footnote{This version is essentially the same as that in Wallace p. 140. Note that Wallace here (i) does not strictly equate the PP with the decision-theoretic link and (ii) sees PP as underpinning not just the decision-theoretic link but also the inferential link (see footnote 24 above). We return to point (i) below; we are also overlooking here the usual subtleties involved with characterising the background information as ``accessible''. For a more careful discussion of the PP, see (Bacciagaluppi 2019).}
 \end{quote}
 
 Now even modest familiarity with the literature reveals that the status of the PP is open to dispute. According to Ned Hall (Hall, 2004), it is an analytic truth (i.e. true by virtue of the meaning of the terms ``chance" and ``credence") though this view is hardly consensual. The weaker position that chance is implicitly defined by the PP has been criticised by Carl Hoefer.\footnote{Hoefer, 2019, section 1.3.4.}
  Amongst other philosophers who think the PP requires justification, Simon Saunders considers that the weakness in the principle has to do with the notoriously obscure nature of chances: 
\begin{quote}
Failing an account of what objective probabilities are, it is hard to see how the [PP] could be justified, for it seems that it ought to be facts about the physical world that dictate our subjective future expectations.(Saunders, 2005)
\end{quote}  
It is precisely the Deutsch-Wallace theorem that Saunders regards as providing the missing account of what chances are, as we shall see. If there is a difference between his view and that of David Wallace, it is that the latter sees the \emph{prima facie} difficulty with the PP as more having to do with the its inferential form than the clarification of one of its components (chances). Wallace argues that justification of the rational inference from chances to credences is generically hard to come by, unless by way of the principles of Everettian physics and decision theory (of which more below).\footnote{Wallace, 2012, sections 4.10, 4.12.}.  In the absence of such principles, Wallace's position seems close to that of Michael Strevens, who in 1999 wrote:
\begin{quotation}
\ldots in order to justify [the PP], it is not enough to simply define objective probability as whatever makes [the PP] rational. In addition, it must be shown that there is something in this world with which it is rational to coordinate subjective probabilities. \ldots

As with Humean induction, so with probability coordination: we cannot conjure a connection between the past and the future, or between the probabilistic and the non-probabilistic, from mathematics and deductive logic alone.\footnote{Strevens, 1999. See also (Hoefer, 2019, sections 1.2.3 and 1.3.4).}
\end{quotation}
Whether the Deutsch-Wallace theorem is the ultimate conjuring trick will be examined shortly. We finish this brief and selective survey of interpretations of the PP with mention of the striking view of Wayne Myrvold, who, like Strevens, wonders how chances could possibly induce normative constraints on our credences. Myrvold cuts the Gordian Knot by claiming that the PP is concerned rather with our \emph{beliefs about chances} -- even if there are no non-trivial chances in the world. 
\begin{quote}
PP captures all we know about credences about chances, since it is not really about chances.\footnote{Myrvold, 'Beyond Chance and Credence' [unpublished manuscript], section 2.5.}
\end{quote}
Can there be any doubt that the meaning and justification of the PP are issues that are far from straightforward?

Subjectivists can of course watch the whole tangled debate with detachment, if not amusement: if there are no chances, nor credences about chances, the PP is empty. And Papineau's two links collapse into one: from frequencies to credences.\footnote{See (Brown, 2011). In this paper it is briefly argued that the related philosophical ``problem of induction'' should be seen as a pseudo-problem.} But now an obvious and old question needs to be addressed: can credences sometimes be determined without any reference to frequencies?

\section{Quantum probability again}

\subsection{The principle of indifference}

It has been recognised from the earliest systematic writings on probability that symmetry considerations can also play an important role in determining credences in chance processes. In fact, in what is often called the ``classical'' theory of probability, which, following Laplace, was dominant amongst mathematicians for nearly a century,\footnote{See (Gillies, 2000, chapter 2) for a useful introduction to the classical theory.} the application of the \emph{principle of indifference} (or insufficient reason) based on symmetry considerations is the grounding of all credences in physics and games of chance. The closely related, so-called ``objective'' Bayesian approach, associated in particular with the name of E. T. Jaynes, likewise makes prominent use of the principle of indifference.

David Wallace voices the opinion of many commentators when he describes this principle as problematic.\footnote{See (Wallace, 2012 section 4.11).} It is certainly stymied in the case of chance processes lacking symmetry (such as involving a biased coin) and it is widely accepted that serious ambiguities arise when the sample space is \emph{continuous}.\footnote{A lucid discussion is found in (Gillies, 2000, pp. 37-48), which contains an insightful critique of Jaynes' defence of the principle of indifference in physics -- and conceding that the principle has been successfully applied in a number of cases in physics.} But what can be wrong with use of the principle when assigning the probability $P(H) = 1/2$ (see section 3.3) in the case of a seemingly unbiased coin? And are we not obtaining in this case a probabilistic credence purely on the basis of a principle of rationality, deriving a `tends to' from a `does', in the words of David Deutsch?

Not according to Wallace. He points to the fact that in the case of a classical (deterministic) chance process, or even a genuinely stochastic process, the symmetry (if any) must ultimately be broken when any one of the possible outcomes is observed. The obvious culprit? Either suitably randomised initial conditions, or stochasticity itself.
\begin{quote}
\dots since only one outcome occurs, something must break the symmetry -- be it actual microconditions of the system, or the actual process that occurs in a stochastic process. Either way, we have to build probabilistic assumptions into the symmetry-breaking process, and in doing so we effectively abandon the goal of explicating probability.\footnote{Wallace, 2012, pp. 147, 148. See also (Wallace, 2010).}
\end{quote}
But if, in a branching Everettian universe where, loosely speaking, everything that can happen does happen, a derivation of credences can be given on the basis of rationality 
principles that feature the symmetry properties of the wavefunction, no such symmetry need be broken, and a reductive analysis of probability is possible. It is precisely this feat that the Deutsch-Wallace theorem achieves.

\subsection{Deutsch}

In 1999, David Deutsch (Deutsch, 1999) attempted to demonstrate two things within the context of Everettian quantum mechanics: (i) the meaning of probabilistic claims in quantum mechanics, which otherwise are ill-defined, and (ii) the fact that the Born Rule need not be considered an extra postulate in the theory. In a more recent paper (Deutsch, 2016), Deutsch stresses that (iii) his 1999 derivation of the Born Rule also overcomes what is commonly called the ``incoherence problem'' in the Everettian account of probabilities.\footnote{This is the problem of understanding how probabilities can come about when everything that can happen does happen. See (Wallace, 2012, pp. 40, 41).}

As for (i), Deutsch interpreted probability as that concept which appears in a Savage-style representation theorem within the application of rational decision theory. In the case of quantum mechanics, Deutsch exploited the fragment of decision theory expunged of any probabilistic notions, applying it to quantum games in an Everettian universe -- or preferences over the future of quantum mechanical measurement outcomes associated with certain payoffs. The emergent notion of probability is agent-dependent, in the sense that, as with Feynman's notion of probability (recall subsection 3.3 above), probabilities arise from the actions of ideally rational agents and have no independent existence in the universe -- a branching universe whose fundamental dynamics is deterministic.
\begin{quote}
\ldots the usual probabilistic terminology of quantum theory is justifiable in light of result of this paper, provided one understands it all as referring ultimately to the behaviour of rational decision makers. \ldots [probabilistic predictions] become implications of a purely factual theory, rather than axioms whose physical meanings are undefined.\footnote{Deutsch, 1999, p.14. As Hemmo and Pitowsky noted, Deutsch's proof, if successful, would give ``strong support to the subjective approaches to probability in general.'' (Hemmo and Pitowsky, 2007, p. 340)}
\end{quote}

What distinguishes the Deutsch argument from the usual decision-theoretic representation theorem in analogous single-world scenarios is that rational agents are constrained to behave not just in conformity with probability theory, but the values of the probabilities are uniquely defined by the state of the system.
It is essential in the 1999 derivation that preferences are based on states which are represented solely by the standard pure and mixed states in quantum mechanics; hidden variables are ruled out. 

Deutsch stressed that the pivotal result concerns the special case where the pure state is a symmetric superposition of two eigenstates of the observable being measured (i.e. the coefficients, or amplitudes, in the superposition are equal). Based on several rules of rationality, Deutsch showed that in this case a rational decision maker behaves as if she believed that each of the two possible outcomes has equal probability, and that she was maximising the probabilistic expectation value of the payoff (expected utility). The equal probability conclusion in this case might be considered a simple consequence of the principle of indifference, but Deutsch is intent on showing by way of decision theory that it makes sense to assign preferences even in the case of ``indeterminate'' outcomes, i.e., Everettian branching (see point (iii) above).

\subsection{Wallace}

Wallace's version of the Deutsch-Wallace (DW) theorem evolved through a series of iterations, starting in 2003 and culminating in his magisterial book on the Everett interpretation of 2012. Wallace attempted to turn Deutsch's ``minimalist'' proof of the Born Rule into a highly rigorous decision-theoretic derivation based on weaker, but more numerous assumptions. 
Unlike Deutsch, Wallace adheres to a dualist interpretation of probability, involving subjective credences and objective chances, and sees both playing a role in the DW-theorem. We spell this out in more detail in section 5.1 below.

What concerns us at this point is the question raised in section 2 whether non-contextualism need be axiomatic in the Everettian picture. In the case of Deutsch's 1999 proof, it is a consequence of (but not equivalent to) an implicit assumption which Wallace was to identify and call \emph{measurement neutrality}; Wallace made it an explicit assumption in his 2003 (Wallace, 2003) reworking of the Deutsch argument. It would be generous to say that a \emph{proof} of non-contextualism obtains in either account.\footnote{Similar qualms were voiced by Hemmo and Pitowsky (Hemmo and Pitowsky, 2007).} The situation is more complicated in Wallace's 2012 proof of the theorem.

In his book, Wallace is at pains to show the role of what he calls non-contextualism. His \emph{non-contextual inference theorem} is the decision theoretic analogue of Gleason's theorem:
\begin{quote}
This theorem states that any solution to a quantum decision problem, provided that the problem is richly structured and satisfies the assumptions of Chapter 5 and that the solution satisfies certain rationality constraints similar to those discussed in Chapter 5, is represented by a density operator iff it is noncontextual. (Wallace (2012), p. 214)
\end{quote}
This is followed by proof of the claim that any solution to a quantum decision problem which is compatible with a state dependent solution must be non-contextual.\footnote{\emph{Op. cit.} p. 215.} 

But it is important to note how Wallace defines this condition. Informally, an agent's preferences conform to a probability rule that is non-contextualist in Wallace's terms if it assigns the same probabilities to the outcomes of a measurement of operator $X$ whether or not a compatible operator $Y$ is measured at the same time.\footnote{Wallace (2012), p. 196.} After giving a more formal decision-theoretic definition, Wallace explicitly admits that this is not exactly the principle used in Gleason's theorem, ``but it embodies essentially the same idea''.\footnote{Wallace (2012), p. 214.} We disagree. Wallace's principle, which for purposes of subsequent discussion we call \emph{weak non-contextualism}, involves a single measurement procedure; its violation means that a rational agent prefers ``a given act to the same (knowably the same act, in fact) under a different description, which violates state supervenience (and, I hope, is obviously irrational).''\footnote{Wallace (2012), p. 197.} But the Gleason-related principle, or \emph{strong non-contextualism}, involves mutually incompatible procedures. Now Wallace appears to be referring to this strong non-contextualism when he writes \begin{quotation}
It is fair to note, though, that just as a non-primitive approach to measurement allows one and the same physical process to count as multiple abstractly construed measurements, it also allows one and the same abstractly construed measurement to be performed by multiple physical processes. It is then a nontrivial fact, and in a sense a physical analogue of noncontextuality, that rational agents are indifferent to which particular process realizes a given measurement.\footnote{\emph{Ibid.}}  
\end{quotation}
But why should rational agents be so indifferent? Because according to Wallace, it is a consequence of the condition of measurement neutrality, which, while having axiomatic status in both Deutsch's 1999 version and Wallace's 2003 version of the DW-theorem, is a trivial corollary of the 2012 Born Rule theorem, which, to repeat, is based on weaker assumptions. It is therefore rationally required. And
\begin{quote}
The short answer as to why is that two acts which correspond to the same abstractly construed measurement can be transformed into the same act via processes to which rational agents are indifferent.\footnote{\emph{Ibid.} Note that a different, recent approach to proving that credences should be non-contextual in quantum mechanics is urged by Steeger (Steeger, J., 'The Sufficient Coherence of Quantum States' [unpublished manuscript]).}
\end{quote}

Now it seems to us, as it did to Pitowsky (see section 2), that to contemplate                                                                                                                                                                                                                                                                                                                                                                                                                                                                                                                                                                                                                                                                                                                                                                                                                                                                                                                                                                                                                                                                                                                                                                                                                                                                                                                                                                                                                                                                                                                                                                                                                                                                                                                                                                                                                                                                                                                                                                                                                                                                                                                                                                                                                                                                                                                                                                                                                                                                                                                                                                                                                                                                                                                                                                                                                                                                                                                                                                                                                                                                                                                                                                                                                                                                                                                                                                                                                                                                                                                                                                                                                                                                                                                                                                                                                                                                                                                                                                                                                                                                                                                                   a \emph{contextual} assignment of probabilities in quantum mechanics is \emph{prima facie} far from  irrational, given the non-commutative property of operators associated with measurements.\footnote{It was mentioned in section 2 above that contextual probabilities may lead to the possibility of superluminal signalling. But this does not imply that contextualism is irrational. Indeed, violation of no-signalling is bound to happen in some ``deviant'' branches in the Everett multiverse; see section 6 below.} In our view, the most plausible justification of non-contextualism in the context of the DW theorem was given by Timpson (Timpson 2011; section 5.1). It is based on consideration of the details of the dynamics of the measurement process in unitary quantum mechanics, and shows that nothing in the standard account of the process supports the possibility of contextualism. However, this argument presupposes that the standard account is independent of the Born Rule, a supposition which deserves attention. At any rate, it should not be overlooked that the DW-theorem applies to systems with Hilbert spaces of arbitrary dimensions, which is a significant advantage over the proof of the Born Rule for credences using Gleason's theorem.

\section{A quantum justification of the Principal Principle?}

\subsection{Wallace and Saunders}

(i) For David Wallace and Simon Saunders, it is of great significance that Everettian quantum mechanics (EQM) provides an underpinning, unprecedented in single-world physics, for the Principal Principle.\footnote{Perhaps this should be understood as an `emancipated' PP, given that Lewis himself did not believe in the existence of chances in a deterministic universe.} For Wallace,
\begin{quote}
\ldots if an Everett-specific derivation of the Principal Principle can be given, then the Everett interpretation solves an important philosophical problem which could not be solved under the assumption that we do not live in a branching universe.\footnote{Wallace, 2012, footnote 26, pp. 150-151.}
\end{quote}
For Saunders, ``\ldots nothing comparable has been achieved for any other physical theory of chance.''\footnote{Saunders, 2010, p. 184.}

And according to Wallace, his latest approach (see below) succeeds where Deutsch's 1999 fails in providing a vindication of the PP. 
\begin{quote}
Deutsch's theorem \ldots amounts simply to a proof of the decision theoretic link between objective probabilities and action.\footnote{Wallace, 2012, p. 237.}
\end{quote}
Clearly, Wallace has a very different reading of the Deutsch theorem to ours; we see no reference to objective probabilities therein. But Wallace is attempting to make a substantive point: that the PP in the context of classical decision theory is not equivalent to Papineau's decision-theoretic link between chances and credences. Here, briefly, is the reason.

The utility function $\cal{U}$ in the Deutsch theorem is, Wallace argues, not obviously the same  as the utility function $\cal{V}$ in classical decision theory, where the agent's degrees of belief may refer to unknown information. $\cal{U}$ is a feature of what Wallace calls the \emph{Minimal decision-theoretic link} between objective probability and action, which supposedly encompasses the Deutsch theorem and Wallace's Born Rule theorem (though see below). In contrast, the standard classical decision theory in which $\cal{V}$ is defined makes no mention of objective probabilities, and allows for bets when the physical state of the system is unknown, and even when there is uncertainty concerning the truth of the relevant physical theory.

For Wallace, the PP will be upheld in Everettian quantum mechanics only when $\cal{U}$ and $\cal{V}$ can be shown to be equivalent (up to a harmless positive affine transformation). If the quantum credences in the Minimal decision-theoretic link are just subjective, or personal probabilities, then the matter is immediately resolved. But in his 2012 book, Wallace feels the need to provide a ``more direct'' proof involving a thought experiment and some further mathematics, resulting in what he calls the \emph{utility equivalence lemma}.\footnote{Wallace, 2012, pp. 208-210 and Appendix D.}  Such considerations are a testament to Wallace's exceptional rigour and attention to detail, but for the skeptic about chances and hence the PP (ourselves included) they seem like a considerable amount of work for no real gain. 

Let's take another look at the Minimal decision theoretic link. The argument is, again, that if preferences over quantum games satisfy certain plausible constraints, the credences defined in the corresponding representation theorem are in turn constrained to agree with branch weights. The arrow of inference goes from suitably constrained credences to (real) numbers extracted from the (complex) state of the system. This is somewhat out of kilt with Papineau's decision-theoretic link, which involves an inference from knowledge of objective probabilities to credences. And this discrepancy is entirely innocuous, we claim, in Deutsch's own understanding of his 1999 theorem, where branch weights are not interpreted as objective, agent-independent chances.  

(ii) As far as we can tell, the grip that the PP has on Wallace's thinking can be traced back to his conviction that probabilities in physics, both in classical statistical mechanics and Everettian quantum mechanics (EQM), represent objective elements of reality, despite the underlying dynamics being fundamentally deterministic. For a philosophical realist to deny the existence of such elements of reality would thus be tantamount to self-refutation (see section 3.1 above). More specifically, a subjective notion of probability, wrote Wallace in 2002,
\begin{quote}
\ldots seems incompatible with the highly objective status played by probability in science in general, and physics in particular. Whilst it is coherent to advocate the abandonment of objective probabilities, it seems implausible: it commits one to believing, for instance, that the predicted decay rate of radioisotopes is purely a matter of belief.\footnote{Wallace, 2002, section 2.7.}
\end{quote}
But subjectivists do \emph{not}  claim that the half-life of $^{14}$C, for example, is purely a matter of belief. It is belief highly constrained by empirical facts. The half-life is arrived at through Bayesian updating based on the results of ever more accurate/plentiful statistical measurements of decay, as we saw in section 3.3.\footnote{In his 2012 book, Wallace does acknowledge this point; see (Wallace, 2012, p. 138).}

For Wallace, in the case of quantum mechanics (and hence of classical statistical mechanics, which he correctly sees as the classical limit of quantum mechanics) the probabilistic elements of reality  -- chances -- are (relative) branch weights, or mod-squared amplitudes. Now no one who is a realist about the quantum state would question whether amplitudes are agent-independent and supervenient on the ontic universal wavefunction. But are they intrinsically ``chances'' of the kind that defenders of the PP would recognise?

This is a hard question to answer, in part because the notion of chance in the literature is so elusive. Wallace and Saunders adopt the approach of ``cautious functionalism''. Essentially, this means that branch weights act as if they were chances, according to the PP. 
Here is Wallace:
\begin{quote}
\ldots the Principal Principle can be used to provide what philosophers call a \emph{functional} definition of objective probability: it defines objective probability to be \emph{whatever thing} fits the `objective probability' slot in the Principal Principle.\footnote{Wallace, 2012, p. 141.} 
\end{quote}
In Saunders' words:
\begin{quote}
[Wallace] shows that branching structures and the squared moduli of the amplitudes, in so far as they are known, \emph{ought} to play the same decision theory role [as in Papineau's decision theoretic link] that chances play, in so far as they are known, in one-world theories.\footnote{Saunders, 2010, p. 184.}
\end{quote}
And again:
\begin{quote}
The [DW] theorem demonstrates that the particular role ordinarily but mysteriously played by physical probabilities, whatever they are, in our rational lives, is played in a \emph{wholly perspicuous and entirely unmysterious way} by branch weights and branching. It is establishing that this role is played by the branch weights, and establishing that they play all the other chance roles, that qualifies these quantities as probabilities.\footnote{See (Saunders, 'The Everett interpretation: probability' [unpublished manuscript]); this paper also contains a rebuttal of the claim that the process of decoherence, so essential to the meaning of branches in EQM, itself depends on probability assumptions.}
\end{quote}

Saunders calls this reasoning a ``derivation'' of the PP\footnote{Saunders, 2010.}, but in our view \emph{it amounts to presupposing the PP and showing that the elusive notion of chance can be cashed out
 in EQM} in terms of branch weights. It hardly seems consistent with Wallace's express hope
 \begin{quote}
 \ldots to find some alternative characterization of objective probability, independent of the Principal Principle, and then prove that the Principal Principle is true for that alternatively characterized notion.\footnote{Wallace, 2012, p. 144.}
  \end{quote}
 Note, however, that Wallace himself stresses that his Born Rule theorem is insufficient for deriving the PP, since it is ``silent on what to do when the quantum state is not known, or indeed when the agent is uncertain about whether quantum mechanics is true." This leads Wallace to think that the theorem, like Deutsch's, merely establishes the \emph{Minimal decision-theoretic link} between objective probability and action, as we have seen. As a consequence Wallace developed a decision-theoretic ``unified approach'' to probabilities in EQM\footnote{\emph{Ibid}, chapter 6}, which avoids the awkwardness of having the concepts of probability and utility derived twice, from his Born Rule theorem, and from appeal to something like the 2010 Greaves-Myrvold de Finetti-inspired solution to what they call the ``evidential problem" in EQM, although a defence of the PP can still be gained by appeal (see above) to the utility equivalence lemma.\footnote{\emph{Ibid}, p. 234.} 
 
 In our view, any such defence relies on the claim that somewhere in the decision-theoretic reasoning an ``agent-independent notion of objective probability"\footnote{\emph{Ibid}, p. 229.} emerges. We turn to this issue now.  
 
(iii) In his 2010 treatment of EQM, Saunders gives an account of ``what probabilities \emph{actually are} (branching structures)''.\footnote{\emph{Ibid}.} 
It is important to recognise the role of the distinction between branching structures and amplitudes in this analysis. The former are `emergent' and non-fundamental; the latter are fundamental and provide (in terms of their modulus squared) the numerical values of the chances:
\begin{quotation}
Just like other examples of reduction, \ldots [probability] can no longer be viewed as fundamental. It can only have the status of the branching structure itself; it is `emergent' \ldots. Chance, like quasiclassicality, is then an `effective' concept, its meaning at the microscopic level entirely derivative on the establishment of correlations, natural or man-made, with macroscopic branching. That doesn't mean that amplitudes in general \ldots have no place in the foundations of EQM -- on the contrary, they are part of the fundamental ontology -- but their link to probability is indirect. It is simply a mistake, if this reduction is successful, to see quantum theory as at bottom a theory of probability.\footnote{Saunders, 2010, p. 182.}
\end{quotation}
Now Everettian subjectivists (about probabilities) concur completely with this conclusion, given that rational agents themselves have the emergent character described in this passage. At any rate, in Saunder's picture, at the microscopic level, amplitudes have nothing to do with chances. Amplitudes only become connected with chances when the appropriate experimental device correlates them with branching structure,\footnote{Simon Saunders, private communication.} and only then, in our view, by way of the PP. 

This notion of chance in EQM is weaker than that often adopted by advocates of objective probability; indeed Wallace himself states that the functional definition of chance has a different character from that of charge, mass or length.\footnote{Wallace, 2012, p. 144. Compare this with the views expressed in section 2.1 above.} The decision-theoretic approach to the Born Rule starts with rational agents and their preferences in relation to quantum games; without the credences emerging from the rationality and structure axioms, application of the PP to infer chances (branch weights) would be impossible. In what sense then are such chances ``out there'', independent of the rational cogitations of agents?

Wallace's answer to this question appears in the Second Interlude of his 2012 book, and in which the Author replies to objections or queries raised by the Sceptic. Here is the relevant passage:
\begin{quote}
SCEPTIC:  Do you really find it acceptable to regard quantum probabilities as defined via decision theory? Shouldn't things like the decay rate of tritium be objective, rather than defined via how much we're prepared to bet on them?

AUTHOR:  The quantum probabilities are objective. In fact, it's clearer in the Everettian context what those objective things are: they're relative branch weights. They'd be unchanged even if there were no bets being made. 

SCEPTIC:  In that case, what's the point of the decision theory?

AUTHOR:  Although branch weights are objective, what makes it true that branch weight=probability is the way in which branch weights figure in the actions of (ideally) rational agents. \ldots\footnote{Wallace, 2012, p. 249. }
\end{quote}

It is easy to see the tension between the Author's replies if one thinks of the objective probabilities emerging from the DW-theorem as having an existence independent of betting agents.\footnote{Wallace, \emph{ibid}, compares branch weights with physical money, in our view a very apt analogy. A dollar note, e.g., has no more intrinsic status as money than branch weights have the status of probability, until humans confer this status on it. This point is nicely brought out by Harari, who writes that the development of money ``involved the creation of a new inter-subjective reality that exists solely in people's shared imagination.'' See Harari, 2014, chapter 10.} But such is not what the application of the PP yields here. And note again that the branch weights yielded by the wavefunction are not objective probabilities (according to the argument) until the branches are effectively defined by the process of decoherence. In an important sense, adoption of something like this derivative notion of chance in the Saunders-Wallace approach is inevitable for chance ``realists'' who defend EQM. If agent-independent chances existed, then there would be more to Everettian ontology than just the universal wavefunction, contrary to the aims of the approach.\footnote{Wallace describes the Everettian program as interpreting the ``bare quantum formalism'' -- which itself makes no reference to probability (Wallace, 2012, p. 16) -- in a ``straightforwardly realist way'' without modifying quantum mechanics (\emph{op. cit.}, p. 36).}

The subjectivist, on the other hand, who views the quantum probabilities as credences, may take (subject to doubts raised in the next section) the DW-theorem to show, remarkably, that there are objective features of the Everettian ``multiverse" that together with rationality principles constrain credences (which are not fundamental elements in EQM) to align with the Born Rule. There is arguably no strict analogue in classical, one-world physics, as Deutsch emphasised. But to call relative branch weights ``objective probabilities'' is a mere \emph{fa\c{c}on de parler}, and the temptation to functionally define them as such only reflects an \emph{a priori} commitment of questionable merit to the validity of the PP.

It is noteworthy that in his 2010 treatment of the DW-theorem, Wallace says the following:
\begin{quote}
The decision-theoretic language in which this paper is written is no doubt necessary to make a properly rigorous case and to respond to those who doubt the very coherence of Everettian probability, but in a way the central core of the argument is not decision-theoretic at all. What is really going on is that the quantum state has certain symmetries and the probabilities are being constrained by those symmetries.\footnote{Wallace, 2010, pp. 259, 260. Saunders has recently also stressed the role of symmetry in the DW theorem; see (Saunders, `The Everett interpretation: probability' [unpublished manuscript], section 6).}
\end{quote}
Note that the probabilities in the argument are credences; the exercise is at heart the application of a sophisticated variant of the principle of indifference based on symmetry. This makes the nature of quantum probabilities essentially something Laplace would recognise, but with the striking new feature mentioned at the beginning of this section. To reify the resulting quantitative values of the credences (branch weights) \emph{as chances} seems to us both unnecessary and ill-advised; it would be like telling Laplace that his credence in the outcome `heads' is a result (in part) of an objective probability inherent in the individual (unbiased) coin.\footnote{Our approach to probability in EQM is very similar to the ``pragmatic'' view of Bacciagaluppi and Ismael, in their thoughtful 2015 review of Wallace's 2012 book (of which more in section 6 below):
\begin{quote}
According to such a view, the ontological content of the theory makes no use of probabilities. There is a story that relates the ontology to the evolution of observables along a family of decoherent histories, and probability is something that plays a role in the cognitive life of an agent whose experience is confined to sampling observables along such a history. In so doing, one would still be doing EQM \ldots (Bacciagaluppi and Ismael, 2015, section 3.2).
\end{quote} 
However, we do not follow these authors in their attempt to define chances consonant with such probabilities and the PP. (We are grateful to David Wallace for drawing to our attention this review paper.)}

\subsection{Earman}

Recently, John Earman (Earman, 2018) has also argued that something like the PP is a ``theorem'' of non-relativistic quantum mechanics, but now in a single-world context. Briefly, the argument goes like this. 

Earman starts the argument within the abstract algebraic formulation of the theory. A "normal" quantum state $\omega$ is defined as a normed positive linear functional on the von Neumann algebra of bounded observables $\cal{B}(\cal{H})$ operating on a separable Hilbert space $\cal{H}$, with a density operator representation. This means that there is a trace class operator $\rho$ on $\cal{H}$ with $Tr(\rho) = 1$, such that $\omega(A) = Tr(\rho A)$ for all $A \in \cal{B}(\cal{H})$. Normal quantum states induce quantum probability functions $Pr^{\omega}$ on the lattice of projections $\cal{P}(\cal{B}(\cal{H}))$:  $Pr^{\omega}(E) = \omega (E) = Tr(\rho E)$ for all $E \in \cal{P}(\cal{B}(\cal{H}))$. \emph{Earman takes the quantum state to be an objective feature of the physical system, and infers that the probabilities induced by them are objective.}\footnote{Op. cit. p. 16.} This is spelt out by showing how pure state preparation can be understood non-probabilistically using the von Neumann projection postulate in the case of yes-no experiments, and using the law of large numbers to relate the subsequent probabilities induced by the prepared state to frequencies in repeated measurements. All this Earman calls the ``top-down'' approach to objective quantum probabilities.

Credences are now introduced into the argument by considering a ``bottom-up'' approach' in which probability functions on $\cal{P}(\cal{B}(\cal{H}))$ are construed as the credence functions of actual or potential Bayesian agents. States are construed as bookkeeping devices used to keep track of credence functions. By considering again a (non-probabilistic) state preparation procedure, and the familiar L\"{u}ders updating rule, Earman argues that a rational agent will in this case, in the light of Gleason's theorem, adopt a credence function on $\cal{P}(\cal{B}(\cal{H}))$ which is precisely the objective probability induced by the prepared pure state (as defined in the previous paragraph).\footnote{Recall that in the typical case of the infinite dimensional, separable Hilbert space $\cal{H}$, a condition of Gleason's theorem is that the probability function must be countably additive; see section 2 above.} Thus, Earman is intent on showing 
\begin{quote}
that there is a straightforward sense in which no new principle of rationality is needed to bring rational credences over quantum events into line with the events' objective chances -- the alignment is guaranteed by as a theorem of quantum probability, assuming the credences satisfy a suitable form of additivity. (Earman, 2018)
\end{quote}

Here are a few remarks on Earman's argument.

\begin{enumerate}

\item Unless the PP in its original guise is interpreted as analytically true, or providing an implicit definition of chance, or, following Myrvold, it is not about real chances at all, the principle is virtually by definition a principle or rule of rationality.\footnote{See, for instance, (Wallace, 2012, section 4.10).} Were it valid, the appeal to Gleason's theorem in Earman's account would be redundant: credences would track the chances supposedly embedded in the algebraic formalism. Earman purports to show that chances and credences in quantum mechanics have separate underpinnings, and that their numerical values coincide when referring to the same measurement outcomes. Rather than providing a quantum theoretical derivation of the PP, Earman is essentially rejecting the PP and attempting to prove a connection between chances and credences whose nature is not that of an extra principle of rationality.

\item The argument in the top-down view is, however, restricted to the algebraic formulation of quantum mechanics that Earman advocates. Other approaches, such as the early de Broglie-Bohm theory and the Everett picture, also take the quantum state -- or that part of it related to the Hilbert space -- to represent a feature of reality. But it is essentially defined therein as a solution to the time (in-)dependent Sch\"{o}dinger equation(s), with no \emph{a priori} connection between the state and probabilities of measurement outcomes, unless in the case of the Everett picture the decisions of agents are taken into account.\footnote{But recall the complication referred to in footnote 64 above.} In the algebraic approach, this connection is baked in, at a very heavy cost: the terms ``measurement'' and ``objective probability'' are primitive.

\item Earman's argument is subject to the same criticism that was raised above in relation to Pitowsky's derivation of the Born Rule: to the extent that they both rely on Gleason's theorem, credence functions are assumed to be not just countably additive but non-contextual. And the derivation breaks down for Hilbert spaces with dimensions less than 3, as Earman recognises.

\end{enumerate}

\section{The refutability issue for Everettian probability}

There is
an important aspect of probability as a \emph{scientific} notion that has been overlooked so far in this paper. It is the falsifiable status of probabilistic claims, as stressed by Karl Popper and more recently Donald Gillies.\footnote{See (Gillies, 2000, chapter 7).} Deutsch himself gives it prominent status in his 2016 paper:
\begin{quote}
If a theory attaches numbers $p_i$ to possible results $a_i$ of an experiment, and calls those numbers `probabilities', and if,
in one or more instances of the experiment, the observed \emph{frequencies} of the $a_i$ differ significantly, according to some [preordained] statistical test, from the $p_i$, then a scientific problem should
be deemed to exist. (Deutsch, 2016)
\end{quote}
In relation to dynamical collapse theories and de Broglie-Bohm pilot wave theory, Deutsch argues that without this methodological rule the probabilities therein are merely decorative, and with it the theories are problematic. The rule, says Deutsch, ``is not a supposed law of nature, nor is it any factual claim about what happens in nature (the explicanda), nor is it derived from one. \ldots And \emph{one cannot make an explanation problematic merely by declaring it so.}'' p. 29. 

For Deutsch, it is a triumph of the DW-theorem that in Everettian quantum mechanics this problem -- that the methodological falsifiability rule is mere whim -- is avoided. The theorem shows that
\begin{quote}
\dots rational gamblers who knew Everettian quantum theory, \ldots and have \ldots made no probabilistic assumptions, when playing games in which randomisers were replaced by quantum measurements, would place their bets as if those were randomisers, i.e. using the [Born] probabilistic rule \dots according to the methodological [falsifiability] rule [above]. p. 31.
\end{quote}
Furthermore, it is argued that ``the experimenter -- who is now aware of the same evidence and theories as they are -- must agree with them'' when one of two gamblers regards his theory as having been refuted.
The argument is subtle, if not convoluted, and requires appeal on the part of gamblers to a non-probabilistic notion of expectation that Deutsch introduces early in the 2016 paper. 

Perhaps a simpler approach works. Subjectivists, as much as objectivists, aim to learn from experience;  despite the immutability of their conditional probabilities (see section 3.3) they of course update their probabilities conditional on new information. As we saw, this is the basis of their version of the law of large numbers. In fact, it might be considered part of the scientific method for subjectivists, or of the very meaning of probabilities in science, that updating is deemed necessary in cases where something like the Popper-Gillies-Deutsch falsifiability rule renders the theory in question problematic. In practice, such behaviour is indistinguishable from that of objectivists about probability in cases of falsification (even in the tentative sense advocated by Deutsch and others). But this line of reasoning does not make special appeal to a branching universe, and thus is not in the spirit of Deutsch's argument.

The reader can decide which, if either, of these approaches is right. But now an interesting issue arises regarding the DW-theorem, as a number of commentators have noticed.

Given the branching created in (inter alia) repeated measurement processes, it is inevitable that in some branches statistics of measurement outcomes will be obtained that fail to adhere to the Born Rule, whatever reasonable preordained statistical test is chosen. These are sometimes called \emph{deviant} branches. What will Everettian-favourable observers in such branches -- conclude?\footnote{In his 1999 paper, Deutsch assumed that agents involved in his argument were initially adherents of the non-probabilistic part of Everettian theory. Wallace was influenced by the work of Greaves and Myrvold (Greaves and Myrvold, 2010) who developed a confirmation theory suitable for branching as well as non-branching universes, but he went on to develop his own confirmation theory as part of what he called a ``unified approach'' to probabilities in EQM (see section 4.1(ii) above). However, Deutsch, in his 2016 paper, follows Popperian philosophy and rejects any notion of theory confirmation, thereby explicitly sidelining the work of Greaves and Myrvold, despite the fact that it contains a falsifiability element as well as confirmation.} If they adhere to the Popper-Gillies-Deutsch falsifiability rule, they \emph{must} conclude that their theory is in trouble. Wallace and Saunders correctly point out that from the point of view of the DW-theorem, they happen just to be unlucky.\footnote{See (Wallace, 2012, p. 196) and (Saunders, 'The Everett interpretation: probability' [unpublished manuscript]). Saunders again correctly points out that in an infinite non-branching universe an analogous situation holds.} But such observers will have no option but to question Everettian theory. To say that statistics \emph{trump} theory, including principles of rationality ultimately based on symmetries,\footnote{Read, 2018, p. 138.} is just to say that the falsifiability rule is embraced, and not to embrace it for subjectivists like Deutsch is to confine the probabilistic component of quantum mechanics to human psychology. 

But what part of the theory is to be questioned? Read has recently argued that it is the non-probabilistic fragment of EQM -- the theory an Everettian accepts before adopting the Born Rule. In this case, observers in deviant branches could question the physical assumptions involved in the DW-theorem (for example, that the agent's probabilities supervene solely on the wavefunction), and thus consider it inapplicable to their circumstances. 

Hemmo and Pitowsky, on the other hand, argued in 2007 (Hemmo and Pitowsky, 2007) that such an observer could reasonably question the rationality axioms that lead to non-contextualism in (either version of) the DW-theorem. Here is Wallace's 2012 reply.
\begin{quote}
\dots it should be clear that the justification used in this chapter is not available to operationalists, for (stripping away the technical detail) the Everettian defence of non-contextuality is that two processes are decision-theoretically equivalent if they have the same effect on the physical state of the system, whatever it is. Since operationalists deny that there \emph{is} a physical state of the system, this route is closed to them.\footnote{Wallace, 2012, p. 226.}
\end{quote}
Now we saw in section 2 (footnote 4) that Pitowsky, in particular, denies the physical reality of the quantum state. But it is unclear to us whether this is relevant to the matter at hand. Wallace seems to be referring to what we called in section 4.3 weak non-contextualism, whereas what Hemmo and Pitowsky have in mind is strong non-contextualism. Wallace proceeds to give a detailed classical decision problem (analogous to the quantum decision problem) which contains a condition of ``classical noncontextuality'' that again is justified if it assumed ``that the state space really is a space of physical states''. But in a classical theory, no analogue exists of non-commutativity and the incompatibility of measurement procedures involved in strong non-contextualism. It seems to us that the concern raised by Hemmo and Pitowsky is immune to Wallace's criticism; it does not depend on adopting a non-``operationalist'' stand in relation to the quantum state, and applies just as much to non-deviant as to deviant branches.\footnote{However, Hemmo and Pitowsky also argue that 
\begin{quote}
\ldots in the many worlds theory, not only Born's rule but any probability rule is meaningless. The only way to solve the problem \ldots is by adding to the many worlds theory some stochastic element. (Hemmo and Pitowsky, 2007, p. 334) 
\end{quote}
This follows from the claim that the shared reality of all of the multiple branches in EQM entails  ``one cannot claim that the quantum probabilities might be inferred (say, as an empirical conjecture) from the observed frequencies.'' (p. 337). This is a variant of the ``incoherence'' objection to probabilities in EQM (see footnote 45 above),  but in our opinion it is \emph{not} obviously incoherent to bet on quantum games in a branching scenario (see (Wallace, 2010)).} But we mentioned in section 3.3 above that a strong plausibility argument for non-contextualism in Wallace's decision-theoretic approach has been provided by Timpson.

In the Copenhagen interpretation of quantum mechanics, quantum bets \emph{are} in effect inferred from observed frequencies: the Born Rule has as its justification nothing more than past statistics. This option is open to the Everettian (though adherents of the DW-theorem hope for more). Indeed, whether they are objectivists or subjectivists concerning probability, their reasoning here would be no different from that described in section 3.3 in relation to the determination of the half-life of $^{14}$C. And this threatens now to make the DW-theorem redundant in non-deviant branches.\footnote{Although in their above-mentioned review, Bacciagaluppi and Ismael do not go quite this far, they do not regard the DW-theorem as necessary for 
``establishing the intelligibility of probabilities'' in EQM. By appealing to standard inductive practice based on past frequencies, they argue:
\begin{quote}
The inputs [frequencies] are solidly objective, but the outputs [probabilities] need not be reified.\ldots In either the classical or the Everettian setting, probability emerges uniformly as a secondary or derived notion, and we can tell an intelligible story about the role it plays in setting credences. (Bacciagaluppi and Ismael, 2015, section 3.2)
\end{quote}.}  After all, if statistics trump theory (see above) this should be the case as much for agents in non-deviant as in deviant branches.

Read, following Richard Dawid and Karim Th\'{e}bault (Dawid and Th\'{e}bault, 2014), has essentially endorsed the redundancy -- or what they call the irrelevancy -- conclusion, with one caveat.
\begin{quote}
\ldots the central case in which DW could have any relevance is the \ldots scenario in which EQM is believed, but the agent in question has no statistical evidence for or against the theory. DW might deliver a tighter link between subjective probabilities and branch weights, and therefore put the justification of PP, and the status of quantum mechanical branch weights as objective probabilities, on firmer footing. However, DW is not necessary to establish the rationality of betting in accordance with Born rule probabilities \emph{tout court}.\footnote{Read, 2018, p. 140. Note that Read, like Dawid and Th\'{e}bault, rests the rationality of the statistics-driven route on the Bayesian confirmation analysis given by Greaves and Myrvold \emph{op. cit.}.}
\end{quote}
While we have doubts above as to whether the DW-theorem does justify the PP, Read is obviously right that the past-statistics-driven route to the Born Rule is not available to an agent ignorant of the statistics! But such an epistemologically-limited agent would be very hard to find in practice.

\section{Conclusions}

The principal motivation for this paper was the elucidation of the difference between the views of David Deutsch and David Wallace as regards the significance of the DW-theorem. Deutsch interprets the probabilities therein solely within the Ramsay-de Finetti-Savage tradition of rational choice, now in the context of quantum games. Wallace defends a dualistic interpretation involving objective probabilities (chances) and credences, and, like Simon Saunders, argues that his more rigorous version of the theorem provides, at long last, a ``derivation'' or ``proof'' of David Lewis' Principal Principle.

We have questioned whether this derivation of the PP is sound. In our view, the Wallace-Saunders argument presupposes the PP and provides within it a number to fill the notorious slot representing chances. For proponents of reified chances in the world, this is, nonetheless, a remarkable result. But it raises awkward questions about the scope of the ontology in EQM, and for subjectivists about probability like Deutsch the concerns with the PP should seem misguided in the first place. Our own inclinations are closer to those of Deutsch, and in key respects mirror those of Bacciagaluppi and Ismael in their discussion of the DW-theorem.

We have also questioned the doubts expressed by Hemmo and Pitowsky as to whether all the rationality assumptions in the DW-theorem are compelling, with particular emphasis on the role of non-contextualism. Finally, by considering the falsifiable nature of probabilities in science, we regard the complaint by Dawid and Th\'{e}bault, and largely endorsed by Read, that the DW-theorem is, for all practical purposes, redundant, a serious challenge to those who endorse the arguments of the authors of the theorem.

\section{Acknowledgements}

The authors are grateful to the editors for the invitation to contribute to this volume. HRB would like to acknowledge the support of the Notre Dame Institute for Advanced Study; this project was started while he was a Residential Fellow during the spring semester of 2018. Stimulating discussions over many years with Simon Saunders and David Wallace are acknowledged. We also thank David Deutsch, Chiara Marletto, James Read and Christopher Timpson for discussions, and Guido Bacciagaluppi, Donald Gillies and Simon Saunders for helpful critical comments on the first draft of this paper. We dedicate this article to the memory of Itamar Pitowsky, who over many years was known to and admired by HRB, and to Donald Gillies, teacher to HRB and guide to the philosophy of probability.

\section{References}

Albrecht, A. and Phillips D.:  Origin of probabilities and their application to the multiverse, Phys. Rev. D, 90, 123514 (2014).

Bacciagaluppi, G. Unscrambling subjective and epistemic probabilities, this volume, and http://philsci-archive.pitt.edu/16393/1/probability\%20paper\%20v4.pdf (2019).

Bacciagaluppi, G. and Ismael, J.: Essay Review: The Emergent Multiverse, Phil Sci. 82, 1-20 (2015).

Barnum, H.: No-signalling-based version of Zurek's derivation of quantum probabilities: A note on ``environment-assisted invariance, entanglement, and probabilities in quantum physics''. quant-ph/0312150 (2003).

B\'{e} M.M. and Chechev V.P.: $^{14}$C - Comments on evaluation of decay data, www.nucleide.org/DDEP\_WG/Nucleides/C-14\_com.pdf (2012). Accessed 01 June 2019.

Bell, J.S.: On the problem of hidden variables in quantum mechanics, Rev. Mod. Phys. 38, 447-452 (1966).

Bricmont, J: Bayes, Boltzmann and Bohm: Probabilities in Physics. In: J. Bricmont, D. D\"{u}rr, M.C. Galavotti, G. Ghirardi, F. Petruccione, and N. Zangi (Eds.), Chance in Physics. Foundations and Perspectives, Springer Verlag, Berlin Heidelberg New York (2001).

Brown, H.R.: Curious and sublime: the connection between uncertainty and probability in physics, Phil. Trans. Roy. Soc. 369, 1-15 (2011). doi:10.1098/rsta.2011.0075

Brown, H.R.: Once and for all; the curious role of probability in the Past Hypothesis. http://philsci-archive.pitt.edu/id/eprint/13008 (2017). Forthcoming in: The Quantum Foundations of Statistical Mechanics, D. Bedingham, O. Maroney and C. Timpson (eds.). Oxford University Press, Oxford (2020).

Brown H.R.: The reality of the wavefunction: old arguments and new, in: Philosophers look at Quantum Mechanics, A. Cordero (ed.), Synthese Library 406, Springer (2019), pp. 63-86. http://philsci-archive.pitt.edu/id/eprint/12978.

Brown H.R. and G. Svetlichny G.: Nonlocality and Gleason's Lemma. Part I Deterministic Theories, Found. Phys. 20, 1379-1387 (1990). 

Busch, P.: Quantum States and Generalized Observables: A Simple Proof of Gleason's Theorem, Phys. Rev. Lett. 91, 120403 (2003).

Dawid, R. and Th\'{e}bault, K:  Against the empirical viability of the Deutsch- 
Wallace-Everett approach to quantum mechanics. Stud. Hist. Phil. Mod. Phys. 47, 55-61 (2014).

de Finetti, Bruno: Foresight: Its Logical Laws, Its Subjective Sources, English translation. In: J.E. Kyburg and H.E. Smokler (eds.), Studies in Subjective Probability, pp. 93-158. John Wiley and Sons, New York (1964).

Deutsch, D.: Quantum theory of probability and decisions. Proc. Roy. Soc. Lon. A455, 3129-3137 (1999).

Deutsch, D.: The logic of experimental tests, particularly of Everettian quantum theory, Stud. Hist. Phil. Mod. Phys. 55, 24-33 (2016).    

Earman, John: The Relation between Credence and Chance: Lewis' ``Principal Principle'' Is a Theorem of Quantum Probability Theory. http://philsci-archive.pitt.edu/14822/

Feynman, R. P., Leighton, R. B., and Matthew Sands, M: The Feynman Lectures on Physics, Vol. 1. Addison-Wesley, Reading, Mass. (1965).

Gillies, D.: The Subjective Theory of Probability, Brit. J. Phil. Sci 23(2),138-157 (1972).

Gillies, D.: An Objective Theory of Probability, Methuen \& Co., London (1973).
 
Gillies, D.: Philosophical Theories of Probability. Routledge, London (2000). 

Galavotti, M. C.: Anti-Realism in the Philosophy of Probability: Bruno de Finetti's Subjectivism. Erkenntnis 31, 239-261 (1989).

Gleason, A.: Measures on the Closed Subspaces of a Hilbert Space, J. Math. Mech. 6, 885-894 (1957).

Greaves, H. and Myrvold, W.: Everett and Evidence. In: Saunders, S., Barrett, J., Kent, A. and Wallace, D. (eds.) Many Worlds? Everett, Quantum Theory, and Reality, pp. 264-304. Oxford University Press, Oxford (2010). http://philsci-archive.pitt.edu/archive/0004222.

Hacking, I.: The Emergence of Probability. Cambridge University Press, Cambridge (1984).

Hall, N.: Two Mistakes about Credence and Chance. Aus. J. Phil. 82, 93-111 (2004).

Handfield, T.: A Philosophical Guide to Chance, Cambridge University Press, Cambridge (2012).

Harari, Y.N.: Sapiens. A Brief History of Humankind. Harvill Secker, Random House, London (2014).

Hemmo, M. and Pitowsky, I.: Quantum probability and many worlds, Stud. Hist. Phil. Mod. Phys. 38, 333-350 (2007).

Hoefer, C.: Chance in the World: a Humean Guide to Objective Chance. Oxford University Press, (Oxford Studies in Philosophy of Science); Oxford, Oxford University Press, (2019). 

Hume, D.: An Enquiry Concerning Human Understanding.  Oxford University Press, Oxford (2008).

Ismael, J.: What Chances Could Not Be, Brit. J. Phil. Sci.
47, No. 1, 79-91 (1996).

Ismael, J.: On Chance (or, Why I am only Half-Humean). In: Current Controversies in Philosophy of Science, Shamik Dasgupta (ed.). Routledge, London (2019, forthcoming).

Jaynes, E.T.: Information Theory and Statistical Mechanics. In: Statistical Physics. 1962 Brandeis Lectures in Theoretical Physics, Volume 3. G.E. Uhlenbeck et al. (eds.), pp.181-218. W.A. Benjamin, New York (1963).

Kochen, S. and Specker, E.P.: The Problem of Hidden Variables in Quantum Mechanics. J. Math. Mech. 17, 59-87 (1967).

Lewis, D.: A Subjectivist's Guide to Objective Chance. In: R. C. Jeffrey (Ed.), Studies in Inductive Logic and Probability, Volume 2, pp. 263-293, University of Caifornia Press (1980). Reprinted in: Lewis, D. Philosophical Papers Volume Two, pp. 83-132. Oxford University Press, Oxford (1986).

Maris J.P, Navr\'{a}til P., Ormand W.E., Nam H. and Dean D.J.: Origin of the Anomalous Long Lifetime of $^{14}$C, Phys. Rev. Lett. 106, 202502 (2011).

Maudlin, T.: What could be objective about probabilities?, Stud. Hist. Phil. Mod. Phys. 38, 275-291 (2007).

McQueen K. J. and Vaidman, L. In defence of the self-location uncertainty account of probability in the many-worlds interpretation, Stud. Hist. Phil. Mod. Phys. 66, 14-23 (2019).

Myrvold, W. C.: Probabilities in Statistical Mechanics, In: Oxford Handbook of Probability and Philosophy, C. Hitchcock and A. H\'{a}jek (Eds.), Oxford University Press (2016). Available at http://philsci-
archive.pitt.edu/9957/.

Page, D.N.: Sensible quantum mechanics: Are probabilities only in the mind? arXiv:gr-qc/950702v1 (1995).

Papineau, D.: Many Minds Are No Worse than One, Brit. J. Phil. Sci.
47(2), 233-241 (1996).

Pattie R.W. Jr. et al.: Measurement of the neutron lifetime using a magneto-gravitational trap and in situ detection, Science 360, 627-632 (2018).

Pitowsky, I.: George Boole's `Conditions of Possible Experience' and the Quantum Puzzle, Brit. J. Phil. Sci.
45, No. 1, 95-125 (1994).

Pitowsky, I. Quantum Mechanics as a Theory of Probability, in W. Demopoulos and I. Pitowsky (eds.), Physical Theory and its Interpretation [italics], Springer (2006); pp. 213-240. arXiv:quant-phys/0510095v1.

Read, J.: In defence of Everettian decision theory, Stud. Hist. Phil. Mod. Phys. 63, 136-140 (2018).

Saunders, S.: What is Probability? In: Quo Vadis Quantum Mechanics?, A. Elitzur, S. Dolev, and N. Kolenda (eds.), pp. 209-238. Springer-Verlag, Berlin (2005).

Saunders, S.: Chance in the Everett Interpretation. In: Saunders, S., Barrett, J., Kent, A. and Wallace, D. (eds.) Many Worlds? Everett, Quantum Theory, and Reality, pp. 181-205. Oxford University Press, Oxford (2010).

Skyrms, B: Pragmatism and Empiricism, Yale University Press, New Haven (1984).

Strevens, M.: Objective Probability as a Guide to the World, Philosophical Studies: An International Journal for Philosophy in the Analytic Tradition 95(3), 243-275 (1999).

Svetlichny, G.: Quantum Formalism with State-Collapse and Superluminal Communication,  Founds. Phys. 28(2), 131-155 (1998).

Timpson, C.: Probabilities in Realist Views of Quantum Mechanics, in Probabilities in Physics, C. Beisbart and S. Hartmann (eds.), Oxford University Press, Oxford (2011); chapter 6.

Tipler, F.J.: Quantum nonlocality does not exist, Proc. Nat. Ac. Sci. 111(31), 11281-11286 (2014).

Vaidman, L. (2019). Derivations of the Born Rule, this volume and http://philsci-archive.pitt.edu/15943/1/BornRule24-4-19.pdf 

Wallace, D.: Quantum Probability and Decision Theory, Revisited, arXiv:quant-ph/0211104v1 (2002)

Wallace, D.: Everettian rationality: defending Deutsch's approach to probability in the Everett interpretation, Stud. Hist. Phil. Mod. Phys. 34, 415-439 (2003).

Wallace, D.: How to Prove the Born Rule. In: Saunders, S., Barrett, J., Kent, A. and Wallace, D. (eds.) Many Worlds? Everett, Quantum Theory, and Reality, pp. 227-263. Oxford University Press, Oxford (2010).

Wallace, D.: The Emergent Universe. Quantum Theory according to the Everett Interpretation, Oxford University Press, Oxford (2012).

Wallace, D.: Probability in Physics: Statistical, Stochastic, Quantum. In: A. Wilson (ed.), Chance and Temporal Asymmetry, pp. 194-220. Oxford University Press, Oxford (2014).  http://philsci-archive.pitt.edu/9815/1/wilson.pdf

Wright, V. J. and Weigert, S.: A Gleason-type theorem for qubits based on mixtures of projective measurements, J. Phys. A: Math. Theor.
52(5), 055301 (2019).

 \end{document}